\documentclass{elsart}

\usepackage{graphicx}
\usepackage{epsfig,rotating}
\usepackage[dvips]{color}

\def\be{\begin{equation}}
\def\ee{\end{equation}}
\def\ba{\begin{eqnarray}}
\def\ea{\end{eqnarray}}

\def\lsim{\raise0.3ex\hbox{$\;<$\kern-0.75em\raise-1.1ex\hbox{$\sim\;$}}}
\def\gsim{\raise0.3ex\hbox{$\;>$\kern-0.75em\raise-1.1ex\hbox{$\sim\;$}}}

\def\theta{\vartheta}

\begin{document}

\begin{frontmatter}
\title{Reconciling the ultra-high energy cosmic ray spectrum with 
       Fermi shock acceleration}

\author[NTNU,MPI]{M.~Kachelrie\ss}
\author[APC,INR]{and D.V.~Semikoz}

\address[NTNU]{Institutt for fysikk, NTNU Trondheim, N--7491 Trondheim,
  Norway} 
\address[MPI]{Max-Planck-Institut f\"ur Physik
  (Werner-Heisenberg-Institut), D--80805 M\"unchen, Germany}
\address[APC]{APC, Coll\`ege de France, 
11, pl. Marcelin Berthelot, Paris 75005, France}
\address[INR]{INR RAS, 60th October Anniversary prospect 7a,
  117312 Moscow, Russia}

\date{\hfill MPP-2005-113}

\begin{abstract}
The energy spectrum of ultra-high energy cosmic rays (UHECR) is usually 
calculated for sources with identical properties. Assuming that all
sources can accelerate UHECR protons to the same extremely high maximal 
energy $E_{\max}> 10^{20}$~eV and have the steeply falling 
injection spectrum $1/E^{2.7}$, one can reproduce the measured cosmic
ray flux above $E\gsim 10^{18}$~eV. We show that relaxing the
assumption of identical sources and using a power-law distribution of
their maximal energy allows one to explain the observed UHECR spectrum
with the injection $1/E^2$ predicted by Fermi shock acceleration.

\begin{small}
\end{small}
\end{abstract}
\end{frontmatter}

\section{Introduction}
The origin and the composition of ultra-high energy cosmic rays (UHECR) 
are still unknown. Chemical composition studies~\cite{chem_AGASA,chem_HiRes} 
of the CR flux of both the AGASA~\cite{AGASA} and HiRes~\cite{HiRes}
experiments 
point to the dominance of protons above $10^{18}$~eV, but depend strongly
on the details of hadronic interaction models used. Another signature for 
extragalactic protons is a dip in the CR flux around $5\times
10^{18}$~eV seen in the experimental data of AGASA, Fly's Eye, 
HiRes and Yakutsk.  This dip may be caused by energy losses of protons due
to $e^+e^-$ pair production on cosmic microwave photons~\cite{HS85,bump88} 
and was interpreted first by the authors of Ref.~\cite{dip} as  
signature for the dominance of extragalactic protons in the CR flux.
Finally, the UHECR proton spectrum should be strongly suppressed
above $E\gsim 5\times 10^{19}$~eV due to pion production on cosmic 
microwave photons, the so called Greisen-Zatsepin-Kuzmin (GZK)
cutoff~\cite{gzk}.   
The evidence for or against the presence of this cutoff in the
experimental data is at present contradictory.

In this work, we assume following Ref.~\cite{dip} that UHECRs
with $E\gsim 10^{18}$~eV are mostly extragalactic protons and compare
the theoretical predictions for their energy spectrum to experimental data.  
Several groups of authors have tried previously to explain the observed
spectral shape of UHECR flux using mainly two different
approaches: In the first one, the ankle is identified with the 
transition from a steep galactic, usually iron-dominated component to
extragalactic protons with injection spectrum between $\sim
1/E^2$ and $1/E^{2.2}$~\cite{HSW86,RSB93,bw,Wibig:2004ye,Allard:2005ha,DeMarco:2005ia}.
In the second approach, the dip is a feature of $e^+ e^-$ pair production and
one is able to fit the UHECR spectrum down to $E\sim 10^{18}$~eV using only
extragalactic protons and an injection spectrum between $1/E^{2.6}$ and 
$1/E^{2.7}$~\cite{dip,MBO,Lemoine:2004uw,Stecker:2004xm}. 
The first possibility is not very
predictive and probably in contradiction to the observed isotropy of
the UHECR flux~\cite{KST}, while for the second one an explanation for
the complex spectrum suggested ad-hoc by the authors of Ref.~\cite{dip} is
missing. Moreover, an injection spectrum $1/E^{\alpha}$ with
$\alpha\approx 2.7$ is considerably steeper than $\alpha\approx 2$
predicted by shock acceleration~\cite{shock}.

A basic ingredient of all these analyses is the assumption that the
sources are identical. In particular, it is assumed  that every source
can accelerate protons to the same maximal energy $E_{\max}$,
typically chosen as $10^{21}$~eV or higher. However, one expects that
$E_{\max}$ differs among the sources and that the number of potential
sources becomes smaller and smaller for larger $E_{\max}$. Therefore
two natural questions to ask are i) can one explain the observed CR
spectrum with non-identical sources? And ii), is in this case a good
fit of the CR spectrum possible with $\alpha\sim 2$ as predicted by
Fermi shock acceleration? 

In this letter, we address these two questions and show that choosing a
power-law distribution $dn/dE_{\max}\propto E_{\max}^{-\beta}$ for
$E_{\max}$ allows one to explain the measured energy spectrum e.g.\ for
$\alpha=2$ with $\beta=1.7$.

\section{Fitting the AGASA and HIRES data}

We assume a continuous distribution of CR sources with constant
comoving density up to the maximal redshift $z_{\max}=2$. The flux of
sources further away is negligible above $10^{18}$~eV, the lowest
energy considered by us. For simplicity, we assume the same luminosity
for all sources. Then UHECRs are generated according to the injection
spectrum  
\be
 \frac{dN}{dE}\propto E^{-\alpha} \,\theta (E_{\max}-E)   \,,
\label{single}
\ee
and are propagated until their energy is below $10^{18}$~eV or they 
reach the Earth. The proton propagation was simulated with the Monte
Carlo code of Ref.~\cite{numerics}.
The maximal energy $E_{\max}$ in Eq.~(\ref{single}) is usually chosen as
$E_{\max}=10^{21}$~eV or larger for identical sources. In this case, the
exact value of  $E_{\max}$ influences only the UHECR flux above $\sim 5\times
10^{19}$~eV, i.e. the strength of the GZK suppression, while the flux
at lower energies is independent from $E_{\max}$.

The shape of the observed energy spectrum  between $E\approx 10^{18}$~eV
and the GZK cutoff can be well explained by the modification of the
power-law injection spectrum through the energy losses of extragalactic
protons due to pion and $e^+e^-$ pair production on cosmic microwave
photons~\cite{bump88,dip}. In particular, one can reproduce both the
HiRes data~\cite{HiRes} and the Akeno/AGASA data (below $10^{20}$~eV) with
$\alpha=2.7$, cf. Figs. 1, 2 and Refs.~\cite{dip,MBO}.
In order to combine the AGASA~\cite{AGASA} with the Akeno~\cite{Akeno}
data in Fig.~2, we have
rescaled systematically the AGASA data 10\% downwards in energy which
is well within the uncertainty of the absolute energy scale of
AGASA~\cite{AGASA}.

The use of a power-law for the injection spectrum of UHECRs is
well-motivated by models of shock acceleration~\cite{shock}. However,
these models predict as exponent typically $\alpha\approx 2.0$--2.2.
Moreover, the maximal acceleration energy of a certain source depends
obviously on parameters that vary from source to source like its
magnetic field strength or its size~\cite{acc,Protheroe:2004rt}. 
Therefore, one expects that $E_{\max}$ varies vastly among different
sources with less and less sources able to accelerate cosmic rays to
the  high-energy end of the spectrum.

\begin{figure}
\epsfig{file=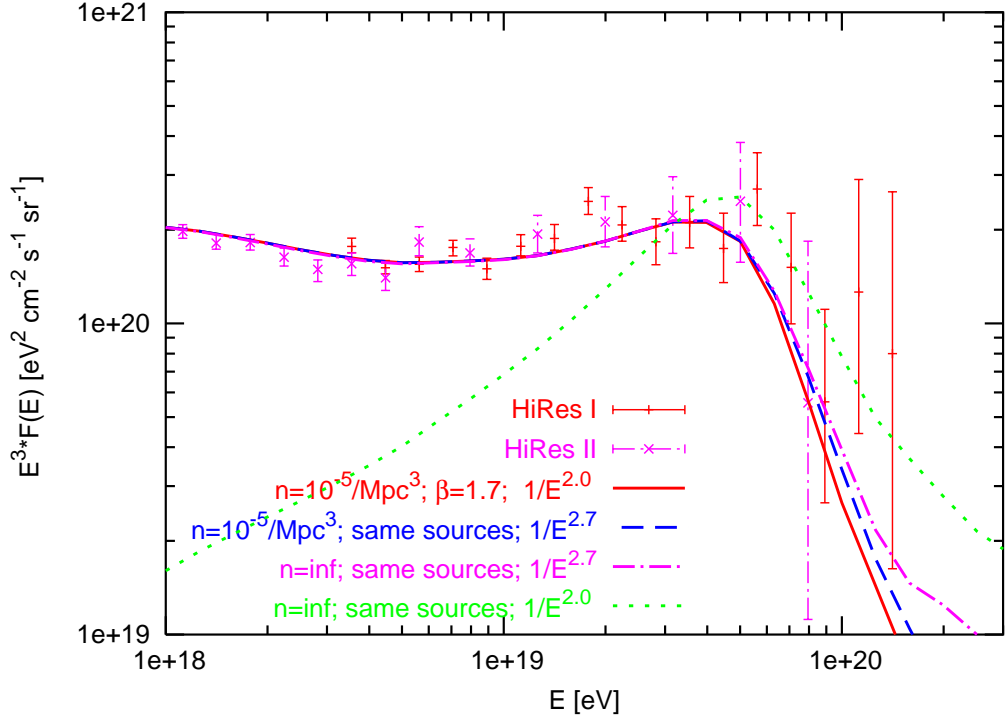,width=0.7\textwidth,angle=270}
\caption{
Fits of the HiRes I and HiRes II data are shown for a uniform
distribution of identical sources with power-law injection spectrum
$1/E^2$ (green, dashed line) and $1/E^{2.7}$ (magenta, dash-dotted
line) for an infinite number of sources as well as for a realistic
source density $n_s= 10^{-5}/{\rm Mpc}^3$ and spectrum $1/E^{2.7}$
(blue, dashed line). The case of an $1/E^2$ spectrum and maximal
energy dependence from Eq.~(\ref{E_max}) with $\beta=1.7$ is shown as
a red, solid line. 
\label{HiRes_fit}}
\end{figure}

\begin{figure}
\epsfig{file=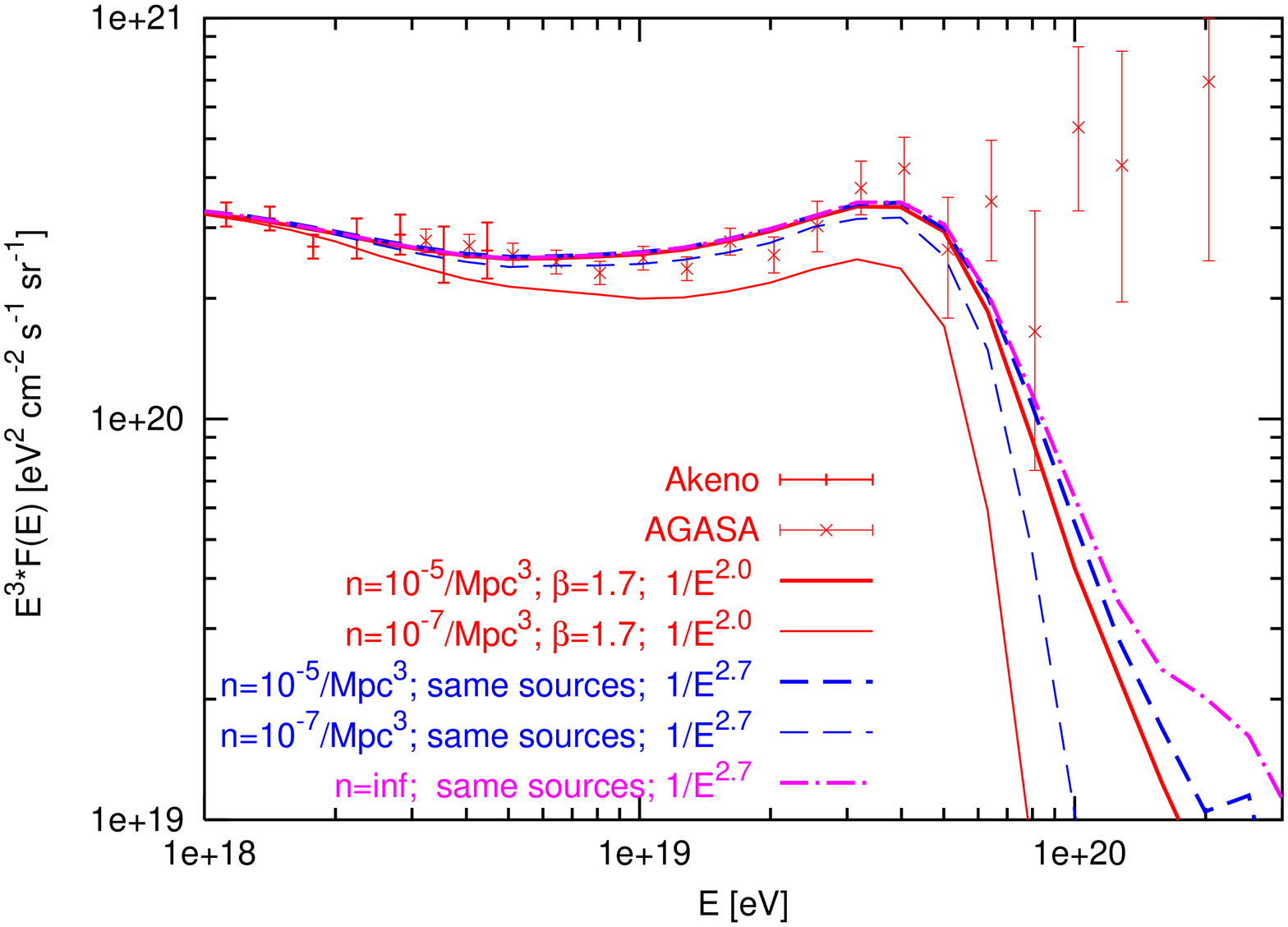,width=0.7\textwidth,angle=270}
\caption{
The fit of  Akeno/AGASA data using a uniform distribution of 
identical sources for an infinite number of sources and power-law spectrum 
$1/E^{2.7}$ is shown as a magenta, dash-dotted line. The same fit with
the realistic source density $n_s=10^{-5}/{\rm Mpc}^3$ and spectrum $1/E^{2.7}$
(thick blue dashed line) and $1/E^2$ spectrum and maximal energy
dependence from Eq.~(\ref{E_max}) with $\beta=1.7$ is shown as a thick
red, solid line. The thin red, solid line for the spectrum $1/E^2$ 
and  $\beta=1.7$  and the thin blue, dashed line for the spectrum $1/E^{2.7}$
correspond to the low source density $n_s=10^{-7}/{\rm Mpc}^3$. 
\label{AGASA_fit}}
\end{figure}

Here, we propose to use more realistic source models for the
calculation of the energy spectrum expected from extragalactic
protons. We relax the assumption of identical sources and suggest to use
a power-law  distribution for the maximal energies of the individual
sources, 
\be
 \frac{dn}{dE_{\max}} \propto E_{\max}^{-\beta}    \,.
\label{E_max}
\ee
Without concrete models for the sources of UHECRs, we cannot derive the
exact form of the distribution of $E_{\max}$ values. However, the use
of a power-law for the $E_{\max}$ distribution is strongly motivated
by the following two reasons: First, we expect a monotonically
decreasing distribution of $E_{\max}$ values and, for the limited
range of two energy decades we consider, a power-law distribution
should be a good approximation to reality. Second, the use of a
power-law distribution for $E_{\max}$ with exponent 
\be
 \beta = \alpha+1-\alpha_0 \,,
\label{index}
\ee
guaranties to recover the spectra calculated with Eq.~(1), i.e.\
$E_{\max}=$~const., for the special case of $E_{\max}\to\infty$ and a
continous distribution of sources\footnote{We are grateful to G.~Sigl
for pointing out this fact.}. 
In Eq.~(\ref{index}), $\alpha$ denotes the exponent of the injection
spectrum of an individual source and $\alpha_0$ the exponent of the
effective injection spectrum after averaging over the $E_{\max}$
distribution of individual sources. For instance, an injection
spectrum of single sources with $\alpha=2$ characteristic for
Fermi shock acceleration reproduces effectively together with a
distribution of $E_{\max}$ values with $\beta=1.7$ the case
$\alpha_0=2.7$ found assuming identical sources.
For finite values of $E_{\max}$ and the source density $n_s$, the
effective injection spectrum is not described anymore by a single
power-law. However, deviations show-up only at energies above $\approx
6\times 10^{19}$~eV or small source densities, see below.

The results for 5.000 Monte Carlo runs of our simulation are
presented in Fig.~\ref{HiRes_fit} for HiRes and in
Fig.~\ref{AGASA_fit} for Akeno/AGASA. In the standard picture of
uniform sources with identical maximal energy (here,
$E_{\max}=10^{21}$~eV) and $1/E^2$ spectrum, 
extragalactic sources contribute only to a few bins of the spectrum
around the GZK cutoff, cf. the green-dotted line in Fig.~\ref{HiRes_fit}.
By contrast, an injection spectrum $1/E^{2.7}$ allows one to
explains the observed data down to $\approx 10^{18}$~eV with
extragalactic protons from identical sources, cf. the magenta, dash-dotted
line for a continuous and the blue, dashed line
for a finite source distribution with $n_s=10^{-5}/{\rm Mpc}^3$ in
Fig.~\ref{HiRes_fit}. This well-known result can be 
obtained also for an injection spectrum $1/E^2$ of individual sources,
if for the $E_{\max}$ distribution, Eq.~(\ref{E_max}), the exponent 
$\beta=1.7$ is chosen. This is illustrated by the red, solid line
in Fig.~\ref{HiRes_fit} for the case of a finite source density
$n_s=10^{-5}/{\rm Mpc}^3$. As already announced, only small differences
at the highest energies, $E\gsim 6\times 10^{19}$~eV, are visible
between an effective $\alpha_0=2.7$ produced by an suitable
$E_{\max}$ distribution and the case of identical sources with
$\alpha=2.7$ for large enough $n_s$.

In Fig.~\ref{AGASA_fit}, we show the dependence of our results
on the source density $n_s$ together with the Akeno/AGASA
data. While for large enough source densities, $n_s=10^{-5}/{\rm Mpc}^3$, 
the spectra  from identical sources with $1/E^{2.7}$ and from
sources with $1/E^2$ injection spectrum, variable $E_{\max}$ and
$\beta=1.7$  are very similar, for smaller densities,
$n=10^{-7}/{\rm Mpc}^3$ in  Fig.~\ref{AGASA_fit}, the shape of the
spectra differs considerably even at lower energies. Thus for small
source densities, the best-fit parameter for $\alpha$ and the
quality of the fit has to be determined for each combination of
$\beta$ and $n_s$ separately and the relation (\ref{index}) is not
valid anymore.

From our results presented in Figs.~\ref{HiRes_fit} and ~\ref{AGASA_fit},   
we conclude that the power-law injection spectrum $1/E^{2.7}$ found
earlier may be seen as a the combined effect of an injection spectrum 
$1/E^2$ predicted by Fermi acceleration and a power-law distribution
of the maximal energies of individual sources with $\beta=1.7$, if the
source density is sufficiently large, $n_s\gsim 10^{-5}/{\rm Mpc}^3$. 
More generally, the exponent $\alpha_0$ obtained from fits assuming
identical sources is connected simply by Eq.~(\ref{index}) to the
parameters $\alpha$ and $\beta$ determining the power-laws of variable
sources in this regime.

\section{Discussion}
The minimal model we proposed can explain the observed UHECR
spectrum for $E>10^{18}$~eV with an injection spectrum as predicted
by Fermi acceleration mechanism, $\alpha=2$--2.2. However, 
in general the experimental data can be fitted for any value of $\alpha$
in the range $2 \le \alpha \le 2.7$ by choosing an appropriate 
index $\beta=\alpha+1-\alpha_0$ in Eq.~(\ref{E_max}).  The best-fit
injection spectrum with $\alpha=2.7$ found for $E_{\max}=$~const.   
appears in our model as an effective value that takes into account the
averaging over the  distribution of $E_{\max}$ values for various sources.

As in the standard case of identical sources, we can not explain the
AGASA excess at  $E>10^{20}$~eV in our model.  Both injection spectra
$1/E^{2.7}$ and $1/E^2$  do not fit well the last three AGASA
bins above $10^{20}$~eV; they have $\chi^2 = 9.8$ and $\chi^2 = 7.9$
respectively. The  shape of the dip can be used also in our model to
understand the overall energy scale of different experiments as
suggested in Ref.~\cite{dip,Berezinsky2005}.

For completeness, we consider now the case of sources with variable
luminosity. The total source luminosity can be defined by
\be
 L(z) = L_0 (1+z)^m  \theta (z_{\max}-z) \theta (z-z_{\min})  \,,
\label{sources}
\ee
where $m$  parameterizes the luminosity evolution, and $z_{\min}$
and $z_{\max}$ are the redshifts of the closest and most distant sources.
Sources in the range $2<z<z_{\max}$ have a negligible contribution 
to the UHECR flux above $10^{18}$~eV. The value of $z_{\min}$ is
connected to the density of sources and influences strongly the shape
of bump and the strength of the GZK suppression~\cite{numerics,density}. 

The value of $m$ influences the spectrum in the range 
$10^{18}~{\rm eV}<E<10^{19}$~eV~\cite{dip}, 
but less strongly than the parameter  
$\beta$ from  Eq.~(\ref{E_max}). Positive values of $m$ increase the
contribution of high-redshift sources and, as a result, injection
spectra with $\alpha<2.7$ can fit the observed data even in the case
of the same $E_{\max}$ for all sources. For example, $\alpha=2.6$ and
$m=3$ fits the AGASA and HiRes data as well as $\alpha=2.7$ and $m=0$
($\chi^2/{\rm d.o.f.}<1$). 
However, a good fit with $\alpha=2$ requires a unrealistic strong
redshift evolution of the sources, $m=16$.

We have presented fits of our model only to the data of Akeno/AGASA and
HiRes. Similar fits can be done for the first results of the
Pierre Auger Observatory~\cite{auger} or the older data of the Yakutsk
experiment~\cite{Yakutsk}. However, the systematic uncertainty of these
data sets is (still) too large, and at present no further insight can
be gained from these data. In the future, data of the Pierre Auger 
Observatory~\cite{auger} and the Telescope Array~\cite{TA} will
restrict the parameter space of theoretical models similar to one
presented here. If a clustered component or even individual sources can be 
identified in the future data, their spectra will allow one to distinguish 
between different possibilities for the injection spectrum. Intriguingly, 
the energy spectrum of the clustered component found by the AGASA 
experiment is much steeper than the overall spectrum~\cite{1}. Thus, one 
might speculate this steeper spectrum is the first evidence for the "true" 
injection spectrum of UHECR sources.

\section{Summary}

In this Letter we have argued that the assumption that all UHECR sources 
accelerate to the same maximal energy is both unrealistic and
unnecessary. Abandoning the idea of identical sources and introducing 
a power-law distribution for the maximal energy of UHECR sources
allows one to fit the CR spectrum above $10^{18}$~eV with the
canonical $1/E^2$ spectrum predicted by Fermi acceleration introducing
$\beta$ as  one additional, physically well-motivated parameter. 
The exponent $\alpha=2.7$ of the best-fit injection spectrum
for identical sources appears in or model only as an effective
parameter, determined by the exponents from the ``true'' injection
spectrum and from the distribution of $E_{\max}$ values.

\section*{Acknowledgments}
We are grateful to Venya Berezinsky, Pasquale Serpico, Igor Tkachev,
Sergey Troitsky and especially to G\"unter Sigl for useful comments.
M.K.\ was partially supported by an Emmy-Noether grant of the DFG.


\end{document}